\documentclass[pra,showpacs,twocolumn]{revtex4}
\usepackage{amssymb}
\usepackage{amsfonts}
\usepackage{amsmath}
\usepackage{graphicx}

\begin{document}

\title{Entropic relations for retrodicted quantum measurements}
\author{Adri\'{a}n A. Budini}
\affiliation{Consejo Nacional de Investigaciones Cient\'{\i}ficas y T\'{e}cnicas
(CONICET), Centro At\'{o}mico Bariloche, Avenida E. Bustillo Km 9.5, (8400)
Bariloche, Argentina, and Universidad Tecnol\'{o}gica Nacional (UTN-FRBA),
Fanny Newbery 111, (8400) Bariloche, Argentina}
\date{\today}

\begin{abstract}
Given an arbitrary measurement over a system of interest, the outcome of a
posterior measurement can be used for improving the statistical estimation
of the system state after the former measurement. Here, we realize an
informational-entropic study of this kind of (Bayesian) retrodicted quantum
measurement formulated in the context of quantum state smoothing. We show
that the (average) entropy of the system state after the retrodicted
measurement (smoothed state) is bounded from below and above by the
entropies of the first measurement when performed in a selective and
non-selective standard predictive ways respectively. For bipartite systems
the same property is also valid for each subsystem. Their mutual
information, in the case of a former single projective measurement, is also
bounded in a similar way. The corresponding inequalities provide a kind of
retrodicted extension of Holevo bound for quantum communication channels.
These results quantify how much information gain is obtained through
retrodicted quantum measurements in quantum state smoothing. While an
entropic reduction is always granted, in bipartite systems mutual
information may be degraded. Relevant physical examples confirm these
features.
\end{abstract}

\maketitle



\section{Introduction}

\textit{Prediction} and \textit{retrodiction} are different and alternative
ways of handling information. Respectively, information in the past or in
the future is taken into account for performing a \textit{probabilistic}
(Bayesian) statement about a system of interest. In physics, most of the
theoretical frames are formulated in a predictive way. The measurement
process in quantum mechanics is clearly predictive. The corresponding
information changes are well known. Non-selective \textit{projective}
measurements never decrease von Neumann entropy \cite{nielsen}. Furthermore,
the entropy $\mathcal{S}[\rho ]\equiv -\mathrm{Tr}[\rho \ln \rho ]$ after a
measurement performed in a \textit{non-selective} way is always greater than
the (average) entropy of the same measurement performed in a \textit{%
selective} way \cite{breuerbook}, that is, $\mathcal{S}[\sum_{k}p(k)\rho
_{k}]\geq \sum_{k}p(k)\mathcal{S}[\rho _{k}],$ where $\rho _{k}$ and $p(k)$
are respectively the system state and probability associated to each outcome 
$k.$ Their difference is bounded by Shannon entropy $\mathcal{H}[k]\equiv
-\sum_{k}p(k)\ln [p(k)]$\ of the outcomes probabilities $\{p(k)\},$ $%
\mathcal{H}[k]\geq \mathcal{S}[\sum_{k}p(k)\rho _{k}]-\sum_{k}p(k)\mathcal{S}%
[\rho _{k}].$ These statements follows straightforwardly from Klein
inequality and the concavity of von Neumann entropy \cite{nielsen,breuerbook}%
. Much less is known when the quantum measurement process is performed in a
retrodictive way.

In quantum mechanics, retrodiction was introduced for criticizing the
apparent time asymmetry of the measurement process \cite{aharonov,vaidman}.
Pre- and post-selected measurement ensembles (initial and final states are
known) are considered. Questions about intermediate states are characterized
through a (retrodictive) Bayesian analysis and the standard Born rule.

Retrodiction also arises in the related formalisms of past quantum states 
\cite{molmer} and quantum state smoothing \cite{wiseman,tsang}, which can be
considered as a quantum extension of classical (Bayesian inference)
smoothing techniques \cite{jaz,recipes}. Both information in the past and in
the future of an open quantum system continuously monitored in time \cite%
{milburn,carmichaelbook} is available. Hence, the system information is
described through a pair of operators, the \textit{past quantum state},
consisting in the system density matrix and an effect operator that takes
into account the future information \cite{molmer}. These objects allow to
estimate the outcome probabilities of an intermediate (retrodicted) quantum
measurement process taking into account both past and future information.
The previous scheme was studied and applied in a wide class of dynamics and
physical arrangements \cite%
{tsanPRA,meschede,murch,haroche,xu,tan,naghi,huard,decay}. The system state
(single density matrix) that takes into account both past and future
information is called \textit{quantum smoothed state} \cite{wiseman,retro}.

While in general it is argued that extra (future) information improves the
estimation of a past (retrodicted) measurement, in contrast with predictive
measurements, a rigorous quantification of this informational benefit is
lacking. Hence, the goal of this paper is to perform an
informational-entropic study of retrodictive quantum measurements. We find
upper and lower bounds for the (average) entropy of the retrodicted state
(quantum smoothed state). They are defined by the entropies of the same
measurement without retrodiction and performed in a non-selective and
selective ways respectively. The same kind of relation is obtained for each
part of a bipartite system. Their mutual information satisfies similar
inequalities whose explicit form (in the case of projective retrodictive
measurements) leads to a kind of retrodicted extension of Holevo bound for
quantum communication channels \cite{nielsen}. These features are
exemplified with a qubit submitted to strong-weak retrodicted measurements 
\cite{murch} and a hybrid quantum-classical optical-like arrange \cite%
{molmer}.

The developed results provide a rigorous characterization of the information
changes achieved through retrodicted quantum measurements. The analysis is
performed in the context of past quantum states and quantum state smoothing 
\cite{molmer,wiseman,tsang}. We remark that retrodicted measurements were
also introduced in alternative ways \cite{barnett,pegg}. Some similitudes
and differences become clear through the present study.

The paper is outlined as follows. In Sec. II we present the general
structure of retrodicted measurements and quantum state smoothing. In Sec.
III the general entropic relations are obtained. The case of bipartite
system is also characterized through their mutual information. In Sec. IV we
study the case of projective measurement performed over a subsystem of a
bipartite arrangement. Retrodicted-like Holevo bounds are derived. Examples
are worked out in Sec. V. In Sec. VI we provide the Conclusions. Calculus
details that support the main results are presented in the Appendices.

\section{Retrodicted quantum measurements}

Here we present the basic scheme (see Fig. 1) corresponding to a retrodicted
quantum measurement. It recovers the past quantum state formalism \cite%
{molmer} and also allows us to define a quantum smoothed state \cite%
{wiseman,retro}.

A quantum system is characterized by its density matrix $\rho _{I}.$ This
object depends on the previous history of the system. In a first step, it is
subjected to an arbitrary measurement process \cite{nielsen,breuerbook}
defined by the set of measurement operators $\{\Omega _{m}\},$ which
fulfills $\sum\nolimits_{m}\Omega _{m}^{\dagger }\Omega _{m}=\mathrm{I,}$
where $\mathrm{I}$ is the identity matrix in the system Hilbert space. The
system states $\{\rho _{m}\}$\ associated to each outcome, and the
probability $\{p(m)\}$\ of their occurrence, respectively are%
\begin{equation}
\rho _{m}=\frac{\Omega _{m}\rho _{I}\Omega _{m}^{\dagger }}{\mathrm{Tr}%
[\Omega _{m}^{\dagger }\Omega _{m}\rho _{I}]},\ \ \ \ \ p(m)=\mathrm{Tr}%
[\Omega _{m}^{\dagger }\Omega _{m}\rho _{I}],  \label{RhoEme}
\end{equation}%
where $\mathrm{Tr}[\bullet ]$ is the trace operation.

After the first measurement, the system evolves with its own (reversible or
irreversible) completely positive dynamics \cite{nielsen,breuerbook} and
then is subjected to a second arbitrary measurement process. It is defined
by a set of operators $\{M_{y}\},$ which satisfy $\sum\nolimits_{y}M_{y}^{%
\dagger }M_{y}=\mathrm{I.}$ In the following analysis the system dynamics is
disregarded, or equivalently, it can be taken into account through a
redefinition of the set of operators $\{M_{y}\}.$

The second measurement implies the state transformation $\rho
_{m}\rightarrow M_{y}\rho _{m}M_{y}^{\dagger }/\mathrm{Tr}[M_{y}^{\dagger
}M_{y}\rho _{m}].$ The (conditional) probability $p(y|m)$ of outcome $y$
given that the first one was $m$ reads%
\begin{equation}
p(y|m)=\mathrm{Tr}[M_{y}^{\dagger }M_{y}\rho _{m}]=\frac{\mathrm{Tr}[\Omega
_{m}\rho _{I}\Omega _{m}^{\dagger }M_{y}^{\dagger }M_{y}]}{\mathrm{Tr}%
[\Omega _{m}^{\dagger }\Omega _{m}\rho _{I}]}.  \label{P_y_cond_m}
\end{equation}

An essential ingredient for defining a retrodicted measurement is to ask
about the inverse conditional probability $p(m|y),$ that is, the probability
of $m$ given the (posterior) outcome $y.$ This object follows from Bayes
rule. Given that the joint probability $p(y,m)$\ for the measurement events $%
m$ and $y$ satisfies $p(y,m)=p(y|m)p(m),$ it reads%
\begin{equation}
p(y,m)=\mathrm{Tr}[\Omega _{m}\rho _{I}\Omega _{m}^{\dagger }M_{y}^{\dagger
}M_{y}].  \label{conjunta}
\end{equation}%
Now, by using that $p(y,m)=p(m|y)p(y),$ where%
\begin{equation}
p(y)=\sum_{m}p(m,y),  \label{py}
\end{equation}%
is the probability of outcome $y,$ we obtain%
\begin{equation}
p(m|y)=\frac{\mathrm{Tr}[\Omega _{m}\rho _{I}\Omega _{m}^{\dagger
}M_{y}^{\dagger }M_{y}]}{\sum_{m^{\prime }}\mathrm{Tr}[\Omega _{m^{\prime
}}\rho _{I}\Omega _{m^{\prime }}^{\dagger }M_{y}^{\dagger }M_{y}]}.
\label{Pmy}
\end{equation}%
This retrodicted probability relies on Bayes rules and standard quantum
measurement theory. It arises in pre- and post-selected ensembles (here
defined by $\rho _{I}$ and the outcome $y)$ \cite{aharonov,vaidman} and also
in the past quantum state formalism (see supplemental material in Ref. \cite%
{molmer}). In fact, $p(m|y)$ can be written in terms of the past quantum
state $\Xi \equiv (\rho ,E)$ where the density and effect operators are $%
\rho =\rho _{I}$ and $E=M_{y}^{\dagger }M_{y}$ respectively.%
\begin{figure}[tbp]
\includegraphics[bb=72 34 665 320,angle=0,width=7.cm]{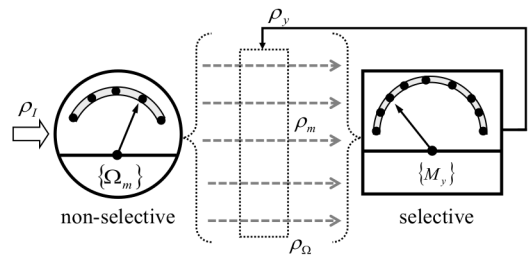}
\caption{Scheme of retrodicted measurements. The system is subjected to two
successive measurement processes defined by the operators $\{\Omega _{m}\}$\
and $\{M_{y}\}$\ respectively. From the second outcome one infers the
probability for the first outcome. The system state at the different stages
is explained in the text.}
\end{figure}

\subsection*{Retrodicted-Quantum smoothed state}

The previous analysis does not associate or define a system state to the
retrodicted probability $p(m|y).$ This assignation depends on extra
assumptions. Similarly to Ref. \cite{molmer} we assume that the result of
the first measurement is hidden to us, that is, the first measurement is a
non-selective one \cite{nielsen,breuerbook}. Hence, the system state after
the first measurement, $\rho _{I}\rightarrow \rho _{\Omega },$ is%
\begin{equation}
\rho _{\Omega }=\sum_{m}\rho _{m}\ p(m)=\sum_{m}\Omega _{m}\rho _{I}\Omega
_{m}^{\dagger }.  \label{Rho_Omega}
\end{equation}

The retrodicted or \textit{smoothed quantum state} $\rho _{y}$ \cite%
{wiseman,retro} here is defined as the estimation of the system state after
the first non-selective measurement \textit{given} that we know the outcome
(labeled by $y$) of the second (selective) measurement. Therefore, we write%
\begin{equation}
\rho _{y}\equiv \sum_{m}\rho _{m}\ p(m|y)=\sum_{m}w(m,y)\Omega _{m}\rho
_{I}\Omega _{m}^{\dagger }.  \label{Rho_y}
\end{equation}%
Here, $w(m,y)\equiv p(m|y)/p(m)=p(y,m)/[p(y)p(m)],$ which from Eqs. (\ref%
{RhoEme}) and (\ref{Pmy}) explicitly reads%
\begin{equation}
w(m,y)=\frac{\mathrm{Tr}[\Omega _{m}\rho _{I}\Omega _{m}^{\dagger
}M_{y}^{\dagger }M_{y}]}{\mathrm{Tr}[\Omega _{m}^{\dagger }\Omega _{m}\rho
_{I}]\sum_{m^{\prime }}\mathrm{Tr}[\Omega _{m^{\prime }}\rho _{I}\Omega
_{m^{\prime }}^{\dagger }M_{y}^{\dagger }M_{y}]}.
\end{equation}

We remark that the smoothed state $\rho _{y}$ depends of (is conditioned to)
the result of the second measurement. Contrarily to the case of pre- and
post selected measurements \cite{aharonov,vaidman}, where $y$ is fixed, here
not any selection is imposed on the second measurement result. Therefore, we
can define an \textit{average smoothed state} $\rho _{M}\equiv \sum_{y}\rho
_{y}\ p(y),$ which corresponds to the system state after averaging $\rho
_{y} $ over the outcomes $y.$ Using that $p(y)=\sum_{m^{\prime }}\mathrm{Tr}%
[\Omega _{m^{\prime }}\rho _{I}\Omega _{m^{\prime }}^{\dagger
}M_{y}^{\dagger }M_{y}]$ [see Eq. (\ref{conjunta})] and that $%
\sum_{y}M_{y}^{\dagger }M_{y}=\mathrm{I},$ it follows%
\begin{equation}
\rho _{M}\equiv \sum_{y}\rho _{y}\ p(y)=\sum_{m}\rho _{m}\ p(m)=\rho
_{\Omega }.  \label{Iguales}
\end{equation}%
Thus, the average smoothed state $\rho _{M}$\ recovers the state $\rho
_{\Omega }$ corresponding to the state after the first non-selective
measurement. A similar property was found in the quantum-classical
arrangements studied in Ref. \cite{retro}.

The analysis of retrodicted quantum measurements performed in Refs. \cite%
{barnett,pegg} also rely on quantum measurement theory and Bayes rule.
Nevertheless, the assumptions are different to the previous ones. After the
second measurement, the state $\rho _{m}$ are not known. Hence, the state
after the first measurement [Eq. (\ref{Rho_Omega})] is taken as a state of
maximal entropy, $\rho _{\Omega }\simeq \mathrm{I,}$ while $\rho _{y}$ [Eq. (%
\ref{Rho_y})] looses its meaning. Hence, the following results do not apply
straightforwardly to those models.

\section{Entropic relations}

The retrodicted quantum measurement scheme described previously consists in
two, non-selective and selective, successive measurements. Now, the relevant
question is how much information gain is obtained from the retrodicted
(smoothed) state $\rho _{y}$ [Eq.(\ref{Rho_y})]. As usual, as an information
measure we consider the von Neumann entropy $\mathcal{S}[\rho ]=-\mathrm{Tr}%
[\rho \ln \rho ].$ In general, one is interested in establishing upper and
lower bounds for $\mathcal{S}[\rho _{y}],$ and to determine how they are
related with, for example, the entropies $\mathcal{S}[\rho _{\Omega }]$ or $%
\mathcal{S}[\rho _{m}].$

Given the arbitrariness of the two measurement processes and given the
random nature of the outcome $y,$ it is not possible to establishing any
general relation between the entropies $\mathcal{S}[\rho _{y}],$ $\mathcal{S}%
[\rho _{\Omega }],$ and $\mathcal{S}[\rho _{m}].$ Any relation is in fact
possible. Therefore, similarly to the case of standard measurement process 
\cite{nielsen,breuerbook}, any entropy relation must be established by
considering averages over the possible measurement outcomes.

By using the concavity of the von Neumann entropy, $\mathcal{S}%
[\sum_{k}p(k)\rho _{k}]\geq \sum_{k}p(k)\mathcal{S}[\rho _{k}]$ \cite%
{nielsen} (with equality if and only if all states $\rho _{k}$ are the
same), in Appendix A we derive the following entropy relation%
\begin{equation}
\mathcal{S}[\rho _{\Omega }]\geq \sum_{y}p(y)\mathcal{S}[\rho _{y}]\geq
\sum_{m}p(m)\mathcal{S}[\rho _{m}].  \label{Central}
\end{equation}%
This is one of the central results of this paper. It demonstrates that the
(average) entropy of the system after the retrodicted measurement, $%
\sum_{y}p(y)\mathcal{S}[\rho _{y}],$ is bounded from above and below by the
entropies of its associated non-selective, $\mathcal{S}[\rho _{\Omega }],$
and (average) selective, $\sum_{m}p(m)\mathcal{S}[\rho _{m}],$ measurement
entropies. In other words, the retrodictive measurement is more informative
than the first non-selective measurement, but is less informative than a
selective resolution of the same measurement process.

In Eq. (\ref{Central}), the lower bound is achieved when all states $\{\rho
_{m}\}$ are the same, or alternatively when $p(m|y)=\delta _{my},$ that is,
both measurement result are completely correlated, $p(y,m)=\delta
_{ym}p(m)=\delta _{my}p(y)$ in Eq.~(\ref{conjunta}). On the other hand, the
upper bound is fulfilled when all states $\{\rho _{y}\}$ are the same. This
last condition occurs when all states $\{\rho _{m}\}$ are identical, or
alternatively when $p(m|y)=p(m).$ Hence, both measurement results, $\{m\}$
and $\{y\},$ are statistically independent, $p(y,m)=p(y)p(m)$ in Eq.~(\ref%
{conjunta}) (see Appendix A).

Interestingly, it is also possible to bound the difference between the terms
appearing in Eq. (\ref{Central}). By using the upper bound $\sum_{k}p(k)%
\mathcal{S}[\rho _{k}]+\mathcal{H}[k]\geq \mathcal{S}[\sum_{k}p(k)\rho _{k}]$
\cite{nielsen}, where $\mathcal{H}[k]=-\sum_{k}p(k)\ln [p(k)]$ is the
Shannon entropy of a probability distribution $\{p(k)\},$ in Appendix A we
obtain%
\begin{equation}
\mathcal{H}[y]\geq \mathcal{S}[\rho _{\Omega }]-\sum_{y}p(y)\mathcal{S}[\rho
_{y}]\geq 0,  \label{HyUpperBound}
\end{equation}%
while in the other extreme it is valid that%
\begin{equation}
\mathcal{H}[m]\geq \sum_{y}p(y)\mathcal{S}[\rho _{y}]-\sum_{m}p(m)\mathcal{S}%
[\rho _{m}]\geq 0.  \label{LowerBound}
\end{equation}%
In this way, the Shannon entropies $\mathcal{H}[y]$ and $\mathcal{H}[m]$
(associated to the two measurement outcomes) bound the difference between
the entropies of the retrodicted and its associated non-selective and
selective measurements. Conditions under which the upper bounds of Eqs. (\ref%
{HyUpperBound}) and (\ref{LowerBound}) are achieved are also provided in
Appendix A.

\subsection{Bipartite systems}

In many physical arrangements where the retrodicted measurement scheme was
studied, the system of interest is a bipartite one. Thus, a relevant
question is to determine if the previous entropy inequality [Eq. (\ref%
{Central})] remains valid (or not) for each subsystem.

Denoting by $A$ and $B$ each subsystem, their states follow from the partial
traces $\rho ^{a}=\mathrm{Tr}_{b}[\rho ^{ab}],$ and $\rho ^{b}=\mathrm{Tr}%
_{a}[\rho ^{ab}],$ where $\rho ^{ab}$ is an arbitrary bipartite state. Under
the replacements $\rho _{m}\rightarrow \rho _{m}^{a/b},$ $\rho
_{y}\rightarrow \rho _{y}^{a/b},$ $\rho _{\Omega }\rightarrow \rho _{\Omega
}^{a/b},$ from the demonstrations of Appendix A it is simple to realize that
the inequalities Eqs.~(\ref{Central}), (\ref{HyUpperBound}), and (\ref%
{LowerBound}) remain valid for each subsystem. This result is valid
independently of which kind of (bipartite) measurements are performed.

\subsection{Mutual information}

Another important aspect that can be studied when considering bipartite
systems is the change in the mutual information between the subsystems. For
a bipartite state $\rho ^{ab},$ the mutual information $\mathcal{I}[\rho
^{ab}]$ is defined as $\mathcal{I}[\rho ^{ab}]\equiv \mathcal{S}[\rho ^{a}]+%
\mathcal{S}[\rho ^{b}]-\mathcal{S}[\rho ^{ab}].$ As demonstrated in Appendix
B, bounds for this object can be derived by using the strong subadditivity
property of von Neumann entropy, $\mathcal{S}[\rho _{abc}]+\mathcal{S}[\rho
_{a}]\leq \mathcal{S}[\rho _{ab}]+\mathcal{S}[\rho _{ac}].$ Thus, as usual
in quantum information results \cite{nielsen}, the demonstrations rely on
introducing an extra ancilla system.

In Appendix B we demonstrate that%
\begin{equation}
\mathcal{S}[\rho _{\Omega }^{ab}]-\sum_{y}p(y)\mathcal{S}[\rho
_{y}^{ab}]\geq \mathcal{I}[\rho _{\Omega }^{ab}]-\sum_{y}p(y)\mathcal{I}%
[\rho _{y}^{ab}].  \label{Mutual_2}
\end{equation}%
Therefore, the difference between the mutual information corresponding to
the non-selective measurement, $\mathcal{I}[\rho _{\Omega }^{ab}],$ and the\
average mutual information corresponding to the retrodicted one, $%
\sum_{y}p(y)\mathcal{I}[\rho _{y}^{ab}],$ is bounded by the positive
quantity $\mathcal{S}[\rho _{\Omega }^{ab}]-\sum_{y}p(y)\mathcal{S}[\rho
_{y}^{ab}]$ [see Eq. (\ref{HyUpperBound})]. On the other hand, based on the
strong subadditivity condition, it is also possible to show that%
\begin{eqnarray}
&&\sum_{y}p(y)\mathcal{S}[\rho _{y}^{ab}]-\sum_{m}p(m)\mathcal{S}[\rho
_{m}^{ab}]\geq  \notag \\
&&\sum_{y}p(y)\mathcal{I}[\rho _{y}^{ab}]-\sum_{m}p(m)\mathcal{I}[\rho
_{m}^{ab}].  \label{Mutual_1}
\end{eqnarray}%
This inequality, which is similar to the previous one, here gives an upper
bound for the difference between the (average) mutual information
corresponding to the retrodicted measurement, $\sum_{y}p(y)\mathcal{I}[\rho
_{y}^{ab}],$ and its (non-retrodicted) selective resolution, $\sum_{m}p(m)%
\mathcal{I}[\rho _{m}^{ab}].$ From Eq. (\ref{LowerBound}) it follows that
the upper bound $\sum_{y}p(y)\mathcal{S}[\rho _{y}^{ab}]-\sum_{m}p(m)%
\mathcal{S}[\rho _{m}^{ab}]$ is a positive quantity.

General conditions under which the previous bounds [Eqs. (\ref{Mutual_2})
and (\ref{Mutual_1})] become equalities are left open \cite{equality}. On
the other hand, notice that only upper bounds were found.

\section{Projective measurements in bipartite systems}

The previous results are general and apply independently of the nature of
the measurement processes (Fig.~1). Here, we consider an arbitrary bipartite
system where a first \textit{projective measurement} is performed on
subsystem $B,$ while the posterior one remains arbitrary being performed
over subsystem $A.$ Hence, the operators $\{\Omega _{m}\}$ that define the
first measurement are written as%
\begin{equation}
\Omega _{m}=\mathrm{I}_{a}\otimes |m\rangle \langle m|.
\end{equation}%
Here, $\mathrm{I}_{a}$ is the identity matrix in the Hilbert space of
subsystem $A$ while $\{|m\rangle \}$ is a complete orthogonal base of $B.$
The second measurement is defined by the set of operators $\{M_{y}\},$ which
act on the Hilbert space of $A.$

The bipartite state associated to each outcome $\{m\}$ reads [Eq.~(\ref%
{RhoEme})]%
\begin{equation}
\rho _{m}^{ab}=\frac{\langle m|\rho _{I}|m\rangle }{\mathrm{Tr}_{a}[\langle
m|\rho _{I}|m\rangle ]}\otimes |m\rangle \langle m|\equiv \rho
_{m}^{a}\otimes |m\rangle \langle m|.  \label{RhoAB_m}
\end{equation}%
The state after the non-selective measurements is [Eq.~(\ref{Rho_Omega})]%
\begin{equation}
\rho _{\Omega }^{ab}=\sum_{m}p(m)\rho _{m}^{a}\otimes |m\rangle \langle m|,
\label{entropiasss}
\end{equation}%
while the retrodictive smoothed state becomes [Eq.~(\ref{Rho_y})]%
\begin{equation}
\rho _{y}^{ab}=\sum_{m}p(m|y)\rho _{m}^{a}\otimes |m\rangle \langle m|.
\label{entropias}
\end{equation}

From Eqs. (\ref{RhoAB_m}) to (\ref{entropias}) it is possible to demonstrate
(Appendix C) that in fact the inequalities (\ref{Central}) and bounds
defined by Eqs. (\ref{HyUpperBound}) and (\ref{LowerBound}) are explicitly
satisfied by the bipartite states. Similar expressions are valid for each
subsystem.

\subsection{Mutual information}

The changes in the mutual information at the different measurement stages
are upper bounded by Eqs. (\ref{Mutual_2}) and (\ref{Mutual_1}). Given the
projective character of the first measurement, here it is also possible to
find a lower bound to these informational changes.

From Eqs. (\ref{Mutual_2}) and Eqs. (\ref{RhoAB_m}) to (\ref{entropias}), in
Appendix C we obtain 
\begin{subequations}
\label{positive2}
\begin{eqnarray}
\mathcal{H}\left[ m:y\right] &\geq &\mathcal{I}[\rho _{\Omega
}^{ab}]-\sum_{y}p(y)\mathcal{I}[\rho _{y}^{ab}] \\
&=&\mathcal{S}[\rho _{\Omega }^{a}]-\sum_{y}p(y)\mathcal{S}[\rho
_{y}^{a}]\geq 0.
\end{eqnarray}%
where $\mathcal{H}[m:y]=\mathcal{H}[m]+\mathcal{H}[y]-\mathcal{H}[m,y],$ is
the classical mutual information between the outcomes of both measurements, $%
\{m\}$ and $\{y\}.$ The lower bound in the previous expression say us that
the (average) mutual information associated to the retrodicted measurement, $%
\sum_{y}p(y)\mathcal{I}[\rho _{y}^{ab}],$ is smaller than that corresponding
to the non-selective measurement, $\mathcal{I}[\rho _{\Omega }^{ab}].$
Hence, contrarily to the entropy measure, here the retrodicted measurement
leads to a degradation of the mutual information between the subsystems.
Similarly to Eq.~(\ref{positive2}), it is possible to obtain (Appendix C) 
\end{subequations}
\begin{subequations}
\label{positive1}
\begin{eqnarray}
\mathcal{H}[m|y] &\geq &\sum_{y}p(y)\mathcal{I}[\rho _{y}^{ab}]-\sum_{m}p(m)%
\mathcal{I}[\rho _{m}^{ab}] \\
&=&\sum_{y}p(y)\mathcal{S}[\rho _{y}^{a}]-\sum_{m}p(m)\mathcal{S}[\rho
_{m}^{a}]\geq 0,\ \ \ \ \ \ \ 
\end{eqnarray}%
where, as before, $\mathcal{H}[m|y]$ is the conditional entropy of outcomes $%
\{m\}$ given outcomes $\{y\}.$ Thus, while the mutual information associated
to the retrodictive measurement decreases with respect to the non-selective
measurement, it is bounded from below by the mutual information of its
selective resolution, $\sum_{m}p(m)\mathcal{I}[\rho _{m}^{ab}].$

\subsection{Retrodicted-like Holevo bound}

Interestingly, Eqs. (\ref{positive2}) and (\ref{positive1}) can be read as a
retrodicted version of the well known Holevo bound for quantum communication
channels \cite{nielsen}.

The standard Holevo bound\ arises in the following context. A sender prepare
a quantum alphabet $\{\rho _{m}^{a}\}$ with probabilities $\{p(m)\}.$ A
receiver performs a measurement characterized by the operators $\{M_{y}\}$
on the sent letter (state), which gives the result $y.$ The Holevo bound
states that for any measurement the receiver may do it is fulfilled that 
\cite{nielsen} 
\end{subequations}
\begin{equation}
\mathcal{H}[m:y]\leq \mathcal{S}\left[ \sum\nolimits_{m}p(m)\rho _{m}^{a}%
\right] -\sum\nolimits_{m}p(m)\mathcal{S}[\rho _{m}^{a}].  \label{Holevo}
\end{equation}%
Hence, the accessible channel information (measured by the mutual
information $\mathcal{H}[m:y]$ between the preparation and the measurement
outcomes), is upper bounded by $\chi \equiv \mathcal{S}\left[
\sum\nolimits_{m}p(m)\rho _{m}^{a}\right] -\sum\nolimits_{m}p(m)\mathcal{S}%
[\rho _{m}^{a}].$

In the retrodicted measurement scheme (Fig. 1), the preparation $\{\rho
_{m}^{a}\}$ with probabilities $\{p(m)\}$ can be associated with the first
non-selective measurement, while the receiver measurement corresponds to the
second one. With this interpretation at hand, we notice that Eq. (\ref%
{positive2}) rewritten as%
\begin{equation}
\mathcal{H}\left[ m:y\right] \geq \mathcal{S}\left[ \sum\nolimits_{y}p(y)%
\rho _{y}^{a}\right] -\sum_{y}p(y)\mathcal{S}[\rho _{y}^{a}],
\label{retroHolevo}
\end{equation}%
can be read as a \textit{retrodicted-like Holevo bound}. While Holevo bound (%
\ref{Holevo}) gives an upper bound for the accessible information, the
retrodicted bound [Eq. (\ref{positive2})] gives a lower bound for $\mathcal{H%
}[m:y].$ Interestingly, it is written in terms of the (retrodicted) quantum
smoothed states $\{\rho _{y}^{a}\}.$ A complementary expression follows
straightforwardly from Eq. (\ref{positive1}).

\section{Examples}

Different experimental realizations of the retrodicted scheme of Fig. 1 are
performed with open quantum systems \textit{continuously monitored in time}.
Hence, their description rely on the formalism of stochastic wave vectors 
\cite{milburn,carmichaelbook}. The results developed in the previous
sections can be extended to this context. Nevertheless, for simplicity, we
consider examples where only two measurements are performed (Fig. 1). The
chosen measurement operators capture the main features of different
experimental realizations \cite{murch,molmer}.

\subsection{Weak and strong retrodicted measurements of a qubit}

First, we consider a qubit system (two-level system) that starts in an
arbitrary state $\rho _{I},$ which is written as%
\begin{equation}
\rho _{I}=\frac{1}{2}(\mathrm{I}+\mathbf{r}_{I}\cdot \mathbf{\sigma }).
\label{RhoInitial}
\end{equation}%
Here, $\mathrm{I}$ is the identity matrix while $\mathbf{\sigma }=(\sigma
_{x},\sigma _{y},\sigma _{z})$ is defined by Pauli matrixes. The Bloch
vector \cite{nielsen,breuerbook} is defined as $\mathbf{r}_{I}=r_{I}\mathbf{n%
},$ where its modulus satisfies $0\leq r_{I}=|\mathbf{r}_{I}|\leq 1$ and $%
\mathbf{n}=(n_{x},n_{y},n_{z})=(\sin (\theta _{I})\cos (\phi _{I}),\sin
(\theta _{I})\sin (\phi _{I}),\cos (\theta _{I})).$ Thus, $\rho _{I}=\rho
_{I}(r_{I},\theta _{I},\phi _{I}).$

Similarly to Ref. \cite{murch}, the first measurement operator is given by $%
(m\rightarrow V)$%
\begin{equation}
\Omega _{V}=(2\pi a^{2})^{-1/4}\exp \left[ -\frac{(V-\sigma _{z})^{2}}{4a^{2}%
}\right] ,  \label{Omega}
\end{equation}%
where $a>0$ is a real free parameter and $V\in (-\infty ,+\infty ),$ which
defines the outcomes of the first measurement. Consistently, $\int_{-\infty
}^{+\infty }dV\Omega _{V}^{\dag }\Omega _{V}=\mathrm{I.}$ In the experiment
analyzed in \cite{murch}, the second measurement can be related with an
effect operator that takes into account the future stochastic evolution.
Instead, here we consider an arbitrary qubit projective measurement $(y=\pm
) $ performed in an arbitrary direction, which is defined by the angles $%
(\theta ,\phi ).$ Hence,%
\begin{equation}
M_{\pm }=|n_{\pm }\rangle \langle n_{\pm }|,  \label{Projector}
\end{equation}%
$[M_{\pm }=M_{\pm }(\theta ,\phi )]$ where the state vectors $|n_{\pm
}\rangle $ are 
\begin{subequations}
\begin{eqnarray}
|n_{+}\rangle &=&\cos \Big{(}\frac{\theta }{2}\Big{)}|+\rangle +\sin \Big{(}%
\frac{\theta }{2}\Big{)}e^{-i\phi }|-\rangle , \\
|n_{-}\rangle &=&-\sin \Big{(}\frac{\theta }{2}\Big{)}|+\rangle +\cos \Big{(}%
\frac{\theta }{2}\Big{)}e^{+i\phi }|-\rangle .
\end{eqnarray}%
Here, $|\pm \rangle $ are the eigenstates of $\sigma _{z}.$

The previous definitions completely set the retrodicted measurement scheme
of Fig. 1. It depends on the free parameters $(r_{I},\theta _{I},\phi
_{I},a,\theta ,\phi ).$\ Our results guarantee that the inequality (\ref%
{Central}) is fulfilled independently of their values.

The states associated to a selective resolution of the first measurement, $%
\rho _{V}=\Omega _{V}\rho _{I}\Omega _{V}^{\dagger }/\mathrm{Tr}[\Omega
_{V}^{\dagger }\Omega _{V}\rho _{I}]$ [Eq.~(\ref{RhoEme})], can be
calculated in an exact way from the following expression 
\end{subequations}
\begin{equation}
\tilde{\rho}_{V}=\sqrt{\frac{1}{2\pi a^{2}}}\left( 
\begin{array}{cc}
\langle +|\rho _{I}|+\rangle e^{-\frac{(V-1)^{2}}{2a^{2}}} & \langle +|\rho
_{I}|-\rangle e^{-\frac{(V^{2}+1)}{2a^{2}}} \\ 
\langle -|\rho _{I}|+\rangle e^{-\frac{(V^{2}+1)}{2a^{2}}} & \langle +|\rho
_{I}|+\rangle e^{-\frac{(V+1)^{2}}{2a^{2}}}%
\end{array}%
\right) ,  \label{RhoVUn}
\end{equation}%
where $\tilde{\rho}_{V}\equiv \Omega _{V}\rho _{I}\Omega _{V}^{\dagger }.$
From this result it is possible to demonstrate that%
\begin{equation}
\lim_{a\rightarrow 0}\rho _{V}=|\pm \rangle \langle \pm |,\ \ \ (V\gtrless
0),\ \ \ \ \ \ \ \lim_{a\rightarrow \infty }\rho _{V}=\rho _{I}.
\label{RhoVlimits}
\end{equation}%
Thus, in the limit $a\rightarrow 0,$ the operators $\{\Omega _{V}\}$ perform
a \textit{strong} projective measurement in the base of eigenstates of $%
\sigma _{z}.$ On the other hand, in the limit $a\rightarrow \infty ,$ a 
\textit{weak} measurement \cite{weak} is performed, $\rho _{V}=\rho _{I}.$

From Eq. (\ref{RhoVUn}), after a straightforward calculation, the state $%
\rho _{\Omega }=\int_{-\infty }^{\infty }dV\ \Omega _{V}\rho _{I}\Omega
_{V}^{\dagger }$ [Eq. (\ref{Rho_Omega})] can be written as%
\begin{equation}
\rho _{\Omega }=\left( 
\begin{array}{cc}
\langle +|\rho _{I}|+\rangle & \langle +|\rho _{I}|-\rangle e^{-\frac{1}{%
2a^{2}}} \\ 
\langle -|\rho _{I}|+\rangle e^{-\frac{1}{2a^{2}}} & \langle +|\rho
_{I}|+\rangle%
\end{array}%
\right) .
\end{equation}%
This expression also reflects the strong and weak feature of the (here
non-selective) measurement as a function of the parameter $a.$ In fact, when 
$a\rightarrow 0$ a diagonal matrix follows, while $\lim_{a\rightarrow \infty
}\rho _{\Omega }=\rho _{I}.$

From Eq. (\ref{RhoVUn}) it is also simple to obtain the probability density $%
p(V)=\mathrm{Tr}[\Omega _{V}^{\dagger }\Omega _{V}\rho _{I}]=\mathrm{Tr}[%
\tilde{\rho}_{V}]$ [Eq.~(\ref{RhoEme})], which is defined by a superposition
of two shifted Gaussian distributions weighted by the initial populations $%
\langle \pm |\rho _{I}|\pm \rangle .$ In general, the joint probability $%
p(\pm ,V)=\mathrm{Tr}[\Omega _{V}\rho _{I}\Omega _{V}^{\dagger }M_{\pm
}^{\dagger }M_{\pm }]$ [Eq.~(\ref{conjunta})] reads%
\begin{eqnarray}
p(\pm ,V) &=&+\sqrt{\frac{1}{2\pi a^{2}}}e^{-\frac{(V\mp 1)^{2}}{2a^{2}}%
}\cos \Big{(}\frac{\theta }{2}\Big{)}^{2}\langle \pm |\rho _{I}|\pm \rangle 
\notag \\
&&+\sqrt{\frac{1}{2\pi a^{2}}}e^{-\frac{(V\pm 1)^{2}}{2a^{2}}}\sin \Big{(}%
\frac{\theta }{2}\Big{)}^{2}\langle \mp |\rho _{I}|\mp \rangle \\
&&\pm \sqrt{\frac{1}{2\pi a^{2}}}e^{-\frac{V^{2}+1}{2a^{2}}}\sin (\theta
)[e^{+i\phi }\langle +|\rho _{I}|-\rangle +c.c.].  \notag
\end{eqnarray}%
From here, it follows the expressions for the retrodicted probabilities $%
p(V|\pm )=\mathrm{Tr}[\Omega _{V}^{\dagger }\Omega _{V}\rho _{I}M_{\pm
}]/p(\pm )$ [Eq.~(\ref{Pmy})], and the probabilities $p(\pm )=\int_{-\infty
}^{+\infty }dV\mathrm{Tr}[\Omega _{V}\rho _{I}\Omega _{V}^{\dagger }M_{\pm
}^{\dagger }M_{\pm }]$ associated to the second measurement outcomes [Eq. (%
\ref{py})]. On the other hand, the integral that define the retrodicted
smoothed states [Eq.~(\ref{Rho_y})]%
\begin{equation}
\rho _{\pm }=\int_{-\infty }^{\infty }dV\ p(V|\pm )\frac{\Omega _{V}\rho
_{I}\Omega _{V}^{\dagger }}{\mathrm{Tr}[\Omega _{V}^{\dagger }\Omega
_{V}\rho _{I}]},  \label{smoothed}
\end{equation}%
must be performed in a numerical way.

\begin{figure}[tbp]
\includegraphics[bb=35 46 735 595,angle=0,width=8.cm]{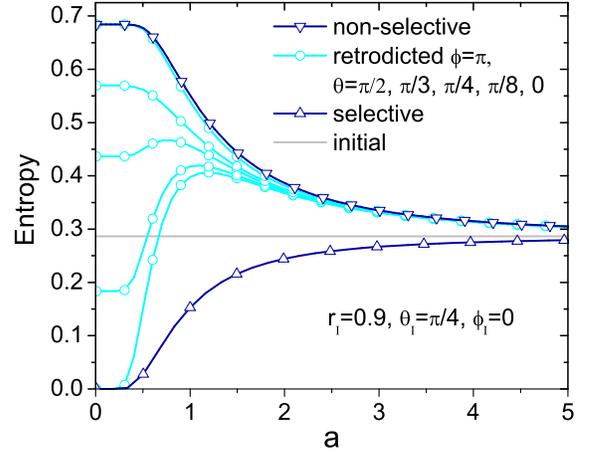}
\caption{Average entropies corresponding to the qubit retrodicted
measurement scheme defined by Eqs. (\protect\ref{Omega}) and (\protect\ref%
{Projector}). The entropies are plotted as a function of the parameter $a$
that defines the first measurement. The angles of the second projective
measurement are $\protect\phi =\protect\pi $ and, from top to bottom, $%
\protect\theta =\protect\pi /2,\protect\pi /3,\protect\pi /8,$ and $0.$ The
parameters of the initial state $(r_{I},\protect\theta _{I},\protect\phi %
_{I})$ [Eq. (\protect\ref{RhoInitial})] are indicated in the plot, jointly
with its entropy (gray line).}
\end{figure}

In Fig. 2, for a particular initial condition, we show the entropies
associated to the nonselective and selective measurements, $\mathcal{S}[\rho
_{\Omega }]$ and $\int_{-\infty }^{+\infty }dVp(V)\mathcal{S}[\rho _{V}]$
respectively, as well as the average entropy of the retrodicted smoothed
state, $\sum_{y=\pm }p(y)\mathcal{S}[\rho _{y}].$ Consistently, we observe
that, independently of the parameter $a$ and angles $(\theta ,\phi )$ that
define the first and second measurements respectively [Eqs. (\ref{Omega})
and (\ref{Projector})], the inequalities (\ref{Central}) are fulfilled.

In the limit $a\rightarrow \infty $ (weak measurement) all entropies
converge to the same value, which is given by the entropy of the initial
state (gray line). In fact, in this limit all states $\rho _{V}$ are the
same [Eq. (\ref{RhoVlimits})], property that guarantees the equality of all
(average) entropies in Eq. (\ref{Central}).

In the limit $a\rightarrow 0$ the first measurement corresponds to a strong
projective one in the basis $\{|\pm \rangle \}$ of eigenstates of $\sigma
_{z}.$ When $\theta =0,$ and arbitrary $\phi ,$ the second projective
measurement is performed in the same basis $\{|\pm \rangle \}.$ Thus, both
measurement outcomes are \textit{completely correlated,} which leads to the
equality of the average entropies of the selective and retrodicted
measurements. On the other hand, for $\theta =\pi /2,$ $\phi =\pi ,$ the
second measurement is performed in the basis of eigenstates of $\sigma _{x}.$
In this case, both measurement outcomes are \textit{statistically independent%
} (see Appendix A), which leads to the equality of the average entropies of
the non-selective and retrodicted measurements. While the previous
properties are strictly fulfilled for $a=0,$ in Fig. 2 they remain
approximately valid for $0\leq a\leq a_{s}\approx 0.4.$ Thus, \textit{from
an entropic point of view}, in that interval the first measurement may be
considered as a projective one. In fact, in all curves of Fig. 2, the value
of the plateau regime around the origin can be estimated by taking into
account two successive projective measurements, the first one being in the $%
z $-direction and the second one in the direction defined by the angles $%
(\theta ,\phi ).$

\subsection*{Post-selected expectation values and entropies}

Under post-selection \cite{murch}, the measurement defined by the operator (%
\ref{Omega}) leads to the so-called weak values \cite{weak}. Here, this
feature is analyzed from an entropic point of view.

From the retrodicted measurement scheme it is possible to define the averages%
\begin{equation}
\langle V_{\Omega }\rangle \equiv \int_{-\infty }^{+\infty }dVVp(V),\ \ \ \
\ \ \ \langle V_{\pm }\rangle \equiv \int_{-\infty }^{+\infty }dVVp(V|\pm ).
\label{weakDef}
\end{equation}%
Here, $\langle V_{\Omega }\rangle $ gives the (unconditional) average of
(the random variable) $V$ associated to the first measurement. On the other
hand, $\langle V_{\pm }\rangle $ is the (conditional) average of $V$ given
that the second measurement outcomes is $y=\pm .$ In agreement with Eq. (\ref%
{Iguales}), they fulfill the relation $\langle V_{\Omega }\rangle
=p(+)\langle V_{+}\rangle +p(-)\langle V_{-}\rangle .$ Furthermore, from
Eq.~(\ref{RhoVUn}) it follows $\langle V_{\Omega }\rangle =\langle +|\rho
_{I}|+\rangle -\langle -|\rho _{I}|-\rangle =\mathrm{Tr}[\rho _{I}\sigma
_{z}]=r_{I}\cos (\theta _{I}).$ Consistently, \textit{anomalous weak values}
are defined by the condition $|\langle V_{\pm }\rangle |>1.$

In Fig. 3(a) and (b) we show the behavior of $\langle V_{\pm }\rangle $ as a
function of the parameter $a.$ As expected, by increasing the parameter $a$
(weak measurement limit) the anomalous property $|\langle V_{\pm }\rangle
|>1 $ may develops. Furthermore, we find that this feature is absent for $%
0\leq a\leq a_{s}\approx 0.4,$ which correspond to the interval where, from
an \textit{entropic point of view}, the first measurement can be
approximated by a strong projective one (plateaus in Fig. 2).

Similarly to expectation values, one can define the \textit{conditional
entropies} $\mathcal{S}[\rho _{\pm }],$ which correspond to the entropies of
each post-selected smoothed state, Eq. (\ref{smoothed}). For the same
parameters values, these objects are shown in Fig. 3(c) and (d). We find
that $\mathcal{S}[\rho _{\pm }]$ do not fulfill the (average) bounds (\ref%
{Central}). In addition, we deduce that this feature cannot be related with
the anomalous property of the weak expectation values. In fact, in general,
any relation may occur, that is, normal or anomalous weak values may develop
while the corresponding conditional entropies may or not be bounded by the
constraints (\ref{Central}).

\subsection{Retrodiction in a bipartite quantum-classical optical-like
hybrid system}

Retrodiction was studied in different physical arranges where the effective
dynamics can be described through a quantum system $(A)$ coupled to
unobservable stochastic classical degrees of freedom $(B)$ \cite{tsang}. The
quantum system is continuously monitored in time. For optical ones, its
fluorescence signal is observed via photon- or homodyne-detection processes 
\cite{wiseman,retro}. In Ref. \cite{molmer}, the state of the (two-state)
classical system randomly modulate the coherent (fluorescent intensity)
system dynamics. In general, one may also consider situations where the
classical subsystem modulate any of the characteristic parameters of the
quantum evolution \cite{sms}. These hybrid dynamics can also be studied from
the present perspective, that is, through the entropic inequality Eq. (\ref%
{Central}) and the mutual information inequalities Eqs. (\ref{positive2})
and (\ref{positive1}).

We consider a hybrid quantum-classical system whose initial bipartite state
is%
\begin{equation}
\rho _{I}^{ab}=\sum_{\mu }q_{\mu }\rho _{\mu }\otimes |c_{\mu }\rangle
\langle c_{\mu }|.  \label{InitialBiparto}
\end{equation}%
Here, $\{\rho _{\mu }\}$ are different states $(\mathrm{Tr}_{a}[\rho _{\mu
}]=1)$ of a quantum two-level system $A,$ while the projectors $\{|c_{\mu
}\rangle \langle c_{\mu }|\}$ represent different (countable) macrostates of
classical system $B.$ Their statistical weights satisfy $\sum_{\mu }q_{\mu
}=1.$ The states $\{\rho _{\mu }\}$ are written as%
\begin{equation}
\rho _{\mu }=\frac{1}{2}(\mathrm{I}+\mathbf{r}_{\mu }\cdot \mathbf{\sigma }),
\label{initialA}
\end{equation}%
where, similarly to Eq. (\ref{RhoInitial}), $\{\mathbf{r}_{\mu }\}$ are
Bloch vectors. Hence, $\rho _{\mu }=\rho _{\mu }(r_{\mu },\theta _{\mu
},\phi _{\mu }).$%
\begin{figure}[tbp]
\includegraphics[bb=35 590 725 1145,angle=0,width=8.5cm]{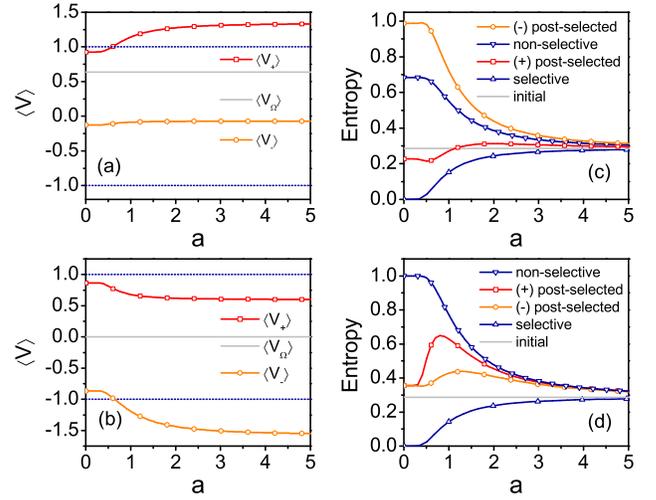}
\caption{(a)-(b) Unconditional and conditional expectations values [Eq. 
\protect\ref{weakDef}] as a function of the parameter $a.$ (c)-(d)
Conditional entropies of the post-selected (smoothed) states [Eq.~(\protect
\ref{smoothed})] jointly with the upper and lower bounds corresponding to
non-selective and selective measurements respectively. In (a)-(c), the
parameters are $(\protect\theta ,\protect\phi )=(\protect\pi /4,\protect\pi )
$ while the initial condition is defined by $(r_{I},\protect\theta _{I},%
\protect\phi _{I})=(0.9,\protect\pi /4,0).$ In (b)-(d), the parameters are $(%
\protect\theta ,\protect\phi )=(\protect\pi /6,0),$ with initial condition $%
(r_{I},\protect\theta _{I},\protect\phi _{I})=(0.9,\protect\pi /2,0).$}
\end{figure}

The first (projective) measurement is defined by the operators $%
(m\rightarrow \mu )$%
\begin{equation}
\Omega _{\mu }=|c_{\mu }\rangle \langle c_{\mu }|,  \label{macro}
\end{equation}%
which are associated to each classical macrostate. The operators of the
second measurement are $(y=\pm )$%
\begin{equation}
M_{+}=|-\rangle \langle +|,\ \ \ \ \ \ \ \ \ M_{-}=|-\rangle \langle -|,
\label{photon}
\end{equation}%
where, as before, $|\pm \rangle $ are the eigenstates of $\sigma _{z},$ and $%
M_{+}^{\dag }M_{+}+M_{-}^{\dag }M_{-}=\mathrm{I.}$ This generalized
measurement \cite{nielsen} can straightforwardly be read as a
photon-detection process. In fact, $M_{+}$ and $M_{-}$ can be associated to
the presence and absence of a transition $|+\rangle \rightsquigarrow
|-\rangle ,$ that is, a photon-detection event. The previous definitions
completely set the retrodicted measurement scheme of Fig.~1.

The state of the bipartite system after a measurement performed with the
operators $\{\Omega _{\mu }\},$ in a selective and non-selective ways,
respectively lead to [Eqs. (\ref{RhoEme}) and (\ref{Rho_Omega})]%
\begin{equation}
\rho _{\mu }^{ab}=\rho _{\mu }\otimes |c_{\mu }\rangle \langle c_{\mu }|,\ \
\ \ \ \ \ \ \ \ \rho _{\Omega }^{ab}=\rho _{I}^{ab}.
\label{ClasicoSelectivoNon}
\end{equation}%
The first expression say us that $\rho _{\mu }$ is the state of $A$ \textit{%
given} that $B$ is in the macrostate $\mu .$ Similarly to the experimental
situations quoted previously, the second equality represent the
inaccessibility of the classical degrees of freedom.

Using that $M_{+}^{\dag }M_{+}=|+\rangle \langle +|$ and $M_{-}^{\dag
}M_{-}=|-\rangle \langle -|,$ the joint probabilities Eq. (\ref{conjunta}) $%
[p(y,m)\rightarrow p(\pm ,\mu )]$ read%
\begin{equation}
p(\pm ,\mu )=q_{\mu }\langle \pm |\rho _{\mu }|\pm \rangle =q_{\mu }\frac{1}{%
2}[1\pm r_{\mu }\cos (\theta _{\mu })].
\end{equation}%
This expression in turn allows us to calculate the retrodicted probabilities 
$\{p(\mu |\pm )\}$ [Eq. (\ref{Pmy})] and $p(\pm )$ [Eq. (\ref{py})]. The
retrodicted smoothed state read [Eq. (\ref{Rho_y})]%
\begin{equation}
\rho _{\pm }^{ab}=\sum_{\mu }p(\mu |\pm )\rho _{\mu }\otimes |c_{\mu
}\rangle \langle c_{\mu }|.
\end{equation}%
Notice that in contrast with projective measurements in arbitrary bipartite
systems [Eq. (\ref{entropias})], here the smoothed state only differs from
the initial condition [Eq. (\ref{InitialBiparto})] by the replacement $%
q_{\mu }\rightarrow p(\mu |\pm ).$ A similar result was found in Ref. \cite%
{retro}.

In order to exemplify the problem we consider a two-state classical system, $%
\mu =1,2.$ Therefore, the free parameters are $(r_{1},\theta _{1},\phi
_{1}), $ $(r_{2},\theta _{2},\phi _{2}),$ for the initial states $\{\rho
_{\mu }\}_{\mu =1,2},$ while an extra parameter $q$ gives their weights in
the initial bipartite state (\ref{InitialBiparto}), $q_{1}=q$ and $%
q_{2}=(1-q).$ Explicit expressions for the entropies and mutual information
can be read from Appendix C.

In Fig. 4(a), for a set of particular initial conditions, we plot the
entropy of the quantum subsystem $A$ as a function of the weight $q.$
Consistently, as demonstrated in Sec.~III, the inequalities (\ref{Central})
are fulfilled by the entropies of the subsystem. In Fig. 4(b) we show the
dependence of the (average) mutual information for the non-selective,
retrodicted and selective measurements schemes. In agreement with Eqs. (\ref%
{positive2}) and (\ref{positive1}), we observe that, while the retrodicted
scheme implies an entropic benefit for each subsystem, the retrodicted
measurement decreases their mutual information when compared with the
non-selective measurement. The difference between these objects is measured
by the retrodicted-like Holevo bound (\ref{retroHolevo}). On the other hand,
the average mutual information for the selective measurement vanishes [see
Eq. (\ref{ClasicoSelectivoNon})]. The main features shown in Fig. 4 remain
valid for arbitrary initial conditions.%
\begin{figure}[tbp]
\includegraphics[bb=50 870 740 1134,angle=0,width=8.5cm]{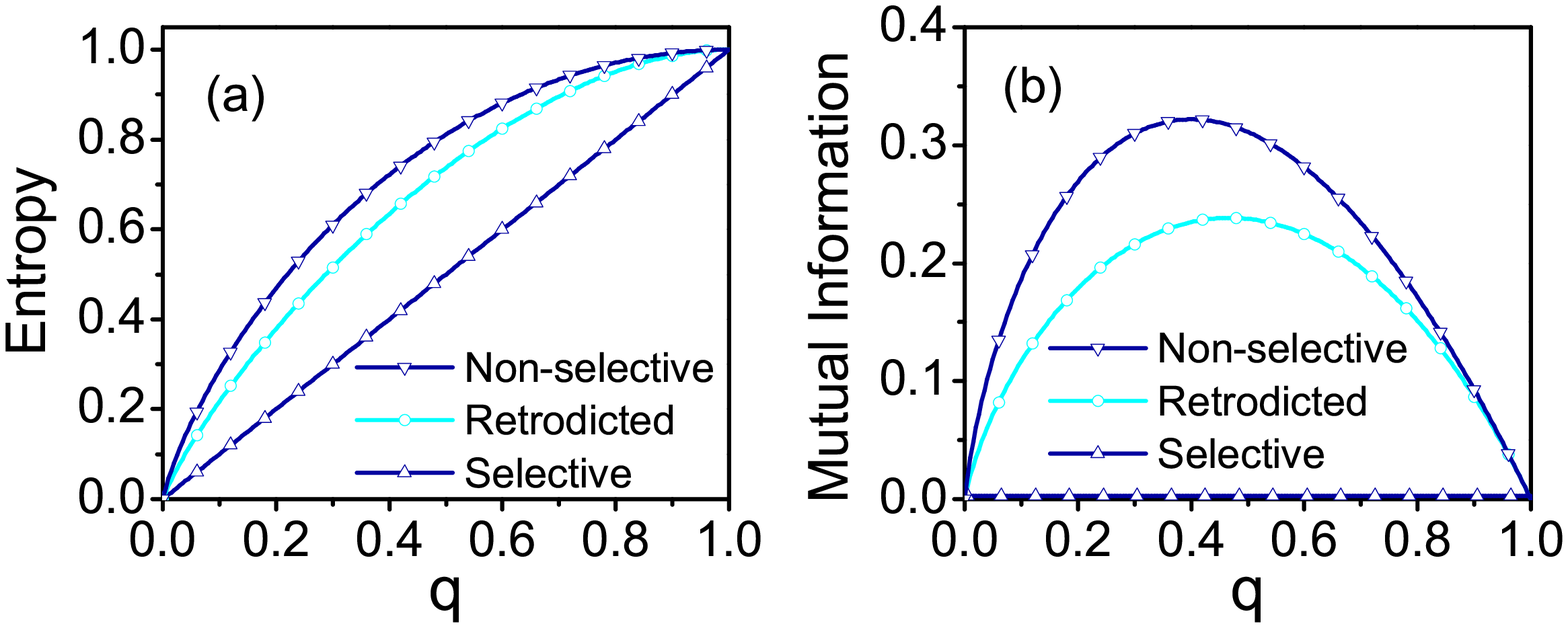}
\caption{(a) Entropy of the quantum subsystem for the measurement scheme
defined by Eqs. (\protect\ref{macro}) and (\protect\ref{photon}). (b) Mutual
information of the quantum-classical arrange. The parameter $q$ define the
weights of the initial bipartite state Eq.~(\protect\ref{InitialBiparto}).
The states of the quantum subsystem $(\protect\rho _{1}$\ and\ $\protect\rho %
_{2})$ [Eq. (\protect\ref{initialA})] are defined with $(r_{1},\protect%
\theta _{1},\protect\phi _{1})=(0,0,0),$ and $(r_{2},\protect\theta _{2},%
\protect\phi _{2})=(1,0,0).$}
\end{figure}

\section{Summary and Conclusions}

We performed and informational-entropic study of retrodicted quantum
measurements (Fig.~1). Given that a non-selective measurement was performed
over a system of interest, a second successive measurement is used for
improving the estimation of the possible outcomes of the former one. From
the quantum expressions for the outcome probabilities, Bayes rule allows to
obtaining the corresponding retrodicted probabilities, Eq.~(\ref{Pmy}). The
system state after the retrodicted measurement (smoothed state) results from
an addition of the system transformations associated to each measurement
outcome with a weight given by the retrodicted probabilities, Eq. (\ref%
{Rho_y}).

Based on the concavity of von Neumann entropy we proved that, in average,
the entropy of the smoothed state is bounded from above and below by the
entropies associated to the first non-selective measurement and the entropy
corresponding to its selective resolution respectively, Eq. (\ref{Central}).
This central result quantifies how much information gain may be obtained
from the retrodicted measurement scheme.

For bipartite systems it was shown that, independently of the measurements
nature, the same property is valid for the entropy of each subsystem. In
addition, based on the strong subadditivity of von Neumann entropy, upper
bounds for the mutual information changes were also established, Eqs. (\ref%
{Mutual_2}) and (\ref{Mutual_1}).

We specified the previous\ results for a bipartite system where the
measurements are performed over each single system successively, being
projective the former one. The retrodicted measurement diminishes the
entropy of each subsystem. Nevertheless, their (average) mutual information
is diminished with respect to that of the non-selective measurement, Eq. (%
\ref{positive2}). This reduction is bounded from below by the (average)
mutual information of the selective resolution of the first measurement, Eq.
(\ref{positive1}). These inequalities in turn lead to a kind of retrodicted
Holevo inequality that bound the (classical) mutual information [Eq. (\ref%
{retroHolevo})] between the two sets of measurement outcomes.

As explicit examples we worked out the case of a qubit subjected to weak and
strong retrodicted measurements. All theoretical results are confirmed by
the model. In addition, we find that anomalous weak values arise when, from
an entropic point of view, the first measurement cannot be approximated by a
strong projective one. On the other hand, we considered a bipartite
quantum-classical optical-like hybrid system. Degradation of mutual
information under the retrodicted measurement scheme was explicitly
confirmed.

The developed results quantify the information changes that follow from a
retrodicted measurement. While the entropy of the system of interest is
always diminished, implying an information vantage, in bipartite systems
mutual information may be degraded. These results provide a solid basis for
studying other informational measures that may be of interest is physical
arrangements where retrodicted measurements are implemented.

\section*{Acknowledgments}

This paper was supported by Consejo Nacional de Investigaciones Cient\'{\i}%
ficas y T\'{e}cnicas (CONICET), Argentina.

\appendix

\section{Demonstration of entropy inequalities}

The entropy inequalities for the retrodicted measurement scheme can be
derived as follows. They rely on the concavity of the von Neumann entropy~%
\cite{nielsen}, $\mathcal{S}[\sum_{k}p(k)\rho _{k}]\geq \sum_{k}p(k)\mathcal{%
S}[\rho _{k}],$ with equality if and only if all states $\rho _{k}$ for
which $p(k)>0$ are identical. Starting from the $\mathcal{S}%
[\sum_{m}p(m)\rho _{m}],$ and using Eq. (\ref{Iguales}), it leads to the
following inequalities 
\begin{subequations}
\begin{eqnarray}
\mathcal{S}\Big{[}\sum_{m}p(m)\rho _{m}\Big{]}\!\! &=&\!\!\mathcal{S}\Big{[}%
\sum_{y}p(y)\rho _{y}\Big{]}\geq \sum_{y}p(y)\mathcal{S}[\rho _{y}],\ \ \ \
\ \ \   \label{Uno} \\
\! &=&\!\sum_{y}p(y)\mathcal{S}\Big{[}\sum_{m}p(m|y)\rho _{m}\Big{]},\ \  \\
\! &\geq &\!\sum_{y}p(y)\sum_{m}p(m|y)\mathcal{S}[\rho _{m}],  \label{Dos} \\
\! &=&\!\sum_{y}\sum_{m}p(m,y)\mathcal{S}[\rho _{m}], \\
\! &=&\!\sum_{m}p(m)\mathcal{S}[\rho _{m}],
\end{eqnarray}%
where we have used that $p(m)=\sum_{y}p(m,y)=\sum_{y}p(m|y)p(y).$ Taking
into account the first and last lines, it follows 
\end{subequations}
\begin{equation}
\mathcal{S}\Big{[}\sum_{m}p(m)\rho _{m}\Big{]}\geq \sum_{y}p(y)\mathcal{S}%
[\rho _{y}]\geq \sum_{m}p(m)\mathcal{S}[\rho _{m}],  \label{EntropyAgain}
\end{equation}%
which recovers the entropy inequalities (\ref{Central}).

Given that \textit{equality} in the concavity entropy inequality is valid if
and only if all states with nonvanishing weight are the same, from Eq.~(\ref%
{Uno}) we deduce that the upper bound is achieved when all states $\{\rho
_{y}\}$ are the same. This condition happens when all states $\{\rho _{m}\}$
are identical, or alternatively when $p(m|y)=p(m)$ [see definition (\ref%
{Rho_y})]. Hence, the joint probability Eq.~(\ref{conjunta}) satisfies $%
p(y,m)=p(y)p(m).$ This condition implies that both measurement results, $%
\{m\}$ and $\{y\},$ are statistically independent. This property is
fulfilled by projective measurements $\Omega _{m}=|m\rangle \langle m|$ and $%
M_{y}=|y\rangle \langle y|,$ where the basis of states $\{|m\rangle \}$ and $%
\{|y\rangle \}$ are such that $|\langle m|y\rangle |^{2}$ is independent of $%
m$ \cite{nota1}.

Similarly, from Eq.~(\ref{Dos}) we deduce that the lower bound in Eq.~(\ref%
{EntropyAgain}) is achieved when all states $\{\rho _{m}\}$ are the same, or
alternatively when $p(m|y)=\delta _{my}.$ Hence, the joint probability Eq.~(%
\ref{conjunta}) satisfies $p(y,m)=\delta _{my}p(y)=\delta _{ym}p(m),$ that
is, both measurement results, $\{m\}$ and $\{y\},$ are completely
correlated. From Eq.~(\ref{conjunta}), we deduce that this condition is
fulfilled by projective measurements $\Omega _{m}=|m\rangle \langle m|$ and $%
M_{y}=|y\rangle \langle y|,$ where the basis of states $\{|m\rangle \}$ and $%
\{|y\rangle \}$ are the same, $|\langle m|y\rangle |^{2}=\delta _{my}.$

We notice that statistical independence and complete correlation between
both measurement outcomes also give the equality conditions for the
entropies of the measurement probabilities $\{p(m)\}$ and their retrodicted
version $\{p(m|y)\}.$ They satisfy the classical inequality \cite{nielsen}%
\begin{equation}
\mathcal{H}[m]\geq \mathcal{H}[m|y]\geq 0,
\end{equation}%
where $\mathcal{H}[m]=-\sum_{m}p(m)\ln [p(m)]$ and $\mathcal{H}%
[m|y]=-\sum_{y}p(y)\sum_{m}p(m|y)\ln [p(m|y)]$ is the conditional Shannon
entropy of outcomes $\{m\}$ given outcomes $\{y\}.$ In fact, $\mathcal{H}[m]=%
\mathcal{H}[m|y]$ when $p(y,m)=p(y)p(m)$ \cite{nielsen}. On the other hand,
the lower bound $\mathcal{H}[m|y]=0$ occurs when $\{m\}$ is a deterministic
function of $\{y\}$ \cite{nielsen}, which here corresponds to $p(y,m)=\delta
_{my}p(y)=\delta _{ym}p(m).$

By using the upper bound \cite{nielsen} $\sum_{k}p(k)\mathcal{S}[\rho _{k}]+%
\mathcal{H}[k]\geq \mathcal{S}[\sum_{k}p(k)\rho _{k}],$ with equality if and
only if all states $\rho _{k}$ have support on orthogonal subspaces, where $%
\mathcal{H}[k]=-\sum_{k}p(k)\ln [p(k)],$ under the replacement $k\rightarrow
y$ it follows%
\begin{equation}
\mathcal{H}[y]\geq \mathcal{S}\Big{[}\sum_{y}p(y)\rho _{y}\Big{]}%
-\sum_{y}p(y)\mathcal{S}[\rho _{y}].
\end{equation}%
Taking into account that $\sum_{y}p(y)\rho _{y}=\sum_{m}p(m)\rho _{m}$ [Eq.~(%
\ref{Iguales})], the previous expression recovers Eq. (\ref{HyUpperBound}).
This upper bound is achieved when all states $\{\rho _{y}\}${}have support
on orthogonal subspaces. On the other hand, taking $k\rightarrow m$ the
upper entropy bound becomes 
\begin{subequations}
\begin{eqnarray}
\mathcal{H}[m] &\geq &\mathcal{S}\Big{[}\sum_{m}p(m)\rho _{m}\Big{]}%
-\sum_{m}p(m)\mathcal{S}[\rho _{m}],\ \ \ \  \\
&\geq &\sum_{y}p(y)\mathcal{S}[\rho _{y}]-\sum_{m}p(m)\mathcal{S}[\rho _{m}],
\end{eqnarray}%
where the last inequality is guaranteed by Eq. (\ref{EntropyAgain}), which
in turn recovers Eq. (\ref{LowerBound}). This upper bound is achieved when
all states $\{\rho _{m}\}$ have support on orthogonal subspaces and $%
p(m|y)=p(m).$

\section{Demonstration of mutual information inequalities}

Here we demonstrate the inequalities that bound the changes in the mutual
information of a bipartite arrangement consisting in subsystems $A$ and $B,$
Eqs. (\ref{Mutual_2}) and (\ref{Mutual_1}). The demonstrations rely on the
strong subadditivity property of von Neumann entropy \cite{nielsen}. Hence,
an extra ancilla system $C$ is introduced.

\textit{First inequality}: The tripartite arrangement is described by the
state 
\end{subequations}
\begin{equation}
\rho ^{abc}=\sum_{m,y}p(m,y)\rho _{m}^{ab}\otimes |y\rangle \langle y|,
\end{equation}%
where $p(m,y)$ is an arbitrary joint probability of $m$ and $y.$ Hence, $%
\sum_{m}p(m,y)=p(y),$ and $\sum_{y}p(m,y)=p(m).$ The set $\{\rho _{m}^{ab}\}$
are states in the $AB$ Hilbert space, $\mathrm{Tr}_{ab}[\rho _{y}^{ab}]=1,$
while $\{|y\rangle \}$ is an orthogonal base of the Hilbert space of $C.$
The marginal state of $AB$ and $C,$ $\rho ^{ab}$ and $\rho ^{c}$\
respectively, read%
\begin{equation}
\rho ^{ab}=\sum_{m}p(m)\rho _{m}^{ab},\ \ \ \ \ \ \ \rho
^{c}=\sum_{y}p(y)|y\rangle \langle y|,
\end{equation}%
where $\rho ^{ab}$ by partial trace defines the states of $A$ and $B,$ $\rho
^{a}=\mathrm{Tr}_{b}[\rho ^{ab}]$ and $\rho ^{b}=\mathrm{Tr}_{a}[\rho ^{ab}]$
respectively. The entropy of the tripartite state $\rho ^{abc},$ by using
that $p(m,y)=p(m|y)p(y),$ can be written as%
\begin{equation}
\mathcal{S}[\rho ^{abc}]=\mathcal{H}[y]+\sum_{y}p(y)\mathcal{S}[\rho
_{y}^{ab}],  \label{Stripartita}
\end{equation}%
where%
\begin{equation}
\rho _{y}^{ab}=\sum_{m}p(m|y)\rho _{m}^{ab},
\end{equation}%
and $\mathcal{H}[y]$ is the classical Shannon entropy of the distribution $%
\{p(y)\},$ $\mathcal{H}[y]=-\sum_{y}p(y)\ln [p(y)].$\ Similarly, the
entropies $\mathcal{S}[\rho ^{ac}]$ and $\mathcal{S}[\rho ^{bc}]$ follows
from Eq. (\ref{Stripartita}) under the replacements $\rho
_{y}^{ab}\rightarrow \rho _{y}^{a}=\mathrm{Tr}_{b}[\rho _{y}^{ab}]$ and $%
\rho _{y}^{ab}\rightarrow \rho _{y}^{b}=\mathrm{Tr}_{a}[\rho _{y}^{ab}]$
respectively. Using the strong subadditivity condition $\mathcal{S}[\rho
^{abc}]+\mathcal{S}[\rho ^{a}]\leq \mathcal{S}[\rho ^{ab}]+\mathcal{S}[\rho
^{ac}]$ \cite{nielsen}, it follows%
\begin{equation}
\mathcal{S}[\rho ^{a}]-\mathcal{S}[\rho ^{ab}]\leq \sum_{y}p(y)\mathcal{S}%
[\rho _{y}^{a}]-\sum_{y}p(y)\mathcal{S}[\rho _{y}^{ab}].
\end{equation}%
Interchanging the indices $a\leftrightarrow b,$ the previous inequality
becomes%
\begin{equation}
\mathcal{S}[\rho ^{b}]-\mathcal{S}[\rho ^{ab}]\leq \sum_{y}p(y)\mathcal{S}%
[\rho _{y}^{b}]-\sum_{y}p(y)\mathcal{S}[\rho _{y}^{ab}].
\end{equation}%
The addition of the previous two expressions lead to%
\begin{equation}
\mathcal{I}[\rho ^{ab}]-\sum_{y}p(y)\mathcal{I}[\rho _{y}^{ab}]\leq \mathcal{%
S}[\rho ^{ab}]-\sum_{y}p(y)\mathcal{S}[\rho _{y}^{ab}],
\end{equation}%
which recovers Eq.~(\ref{Mutual_2}), where the mutual information of a
bipartite state is $\mathcal{I}[\rho ^{ab}]=\mathcal{S}[\rho ^{a}]+\mathcal{S%
}[\rho ^{b}]-\mathcal{S}[\rho ^{ab}].$

\textit{Second inequality}: In this case the tripartite arrangement is
described by the state%
\begin{equation}
\rho _{y}^{abc}=\sum_{m}p(m|y)\rho _{m}^{ab}\otimes |m\rangle \langle m|.
\label{tripartitaY}
\end{equation}%
This state parametrically depends on $y.$ $p(m|y)$ is an arbitrary
conditional probability of $m$ given $y,$ $\sum_{m}p(m|y)=1.$ The set $%
\{\rho _{m}^{ab}\}$ are states in the Hilbert space of the bipartite system $%
AB,$ $\mathrm{Tr}_{ab}[\rho _{y}^{ab}]=1,$ while here $\{|m\rangle \}$ is an
orthogonal base of the Hilbert space of $C.$ The marginal state of $AB$ and $%
C,$ $\rho _{y}^{ab}$ and $\rho _{y}^{c}$\ respectively, read%
\begin{equation}
\rho _{y}^{ab}=\sum_{m}p(m|y)\rho _{m}^{ab},\ \ \ \ \ \ \rho
_{y}^{c}=\sum_{m}p(m|y)|m\rangle \langle m|.
\end{equation}%
The states of $A$ and $B$ read $\rho _{y}^{a}=\mathrm{Tr}_{b}[\rho
_{y}^{ab}] $ and $\rho _{y}^{b}=\mathrm{Tr}_{a}[\rho _{y}^{ab}]$
respectively.

A straightforward calculation leads to%
\begin{equation}
\mathcal{S}[\rho _{y}^{abc}]=\mathcal{H}[m]|_{y}+\sum_{m}p(m|y)\mathcal{S}%
[\rho _{m}^{ab}],  \label{StripartitaY}
\end{equation}%
where%
\begin{equation}
\mathcal{H}[m]|_{y}\equiv -\sum_{m}p(m|y)\ln [p(m|y)].
\end{equation}%
The entropies $\mathcal{S}[\rho _{y}^{ac}]$ and $\mathcal{S}[\rho _{y}^{bc}]$
follows from Eq. (\ref{StripartitaY}) under the replacements $\rho
_{m}^{ab}\rightarrow \rho _{m}^{a}$ and $\rho _{m}^{ab}\rightarrow \rho
_{m}^{b}$ respectively.

Using the strong subadditivity condition \cite{nielsen} $\mathcal{S}[\rho
^{abc}]+\mathcal{S}[\rho ^{a}]\leq \mathcal{S}[\rho ^{ab}]+\mathcal{S}[\rho
^{ac}],$ with $\rho ^{abc}\rightarrow \rho _{y}^{abc}$ [Eq. (\ref%
{tripartitaY})], jointly with Eq. (\ref{StripartitaY}), lead to%
\begin{equation}
\mathcal{S}[\rho _{y}^{a}]-\mathcal{S}[\rho _{y}^{ab}]\leq \sum_{m}p(m|y)%
\mathcal{S}[\rho _{m}^{a}]-\sum_{m}p(m|y)\mathcal{S}[\rho _{m}^{ab}].
\end{equation}%
Interchanging $a\leftrightarrow b$ in the strong subadditivity condition,
the previous equation becomes%
\begin{equation}
\mathcal{S}[\rho _{y}^{b}]-\mathcal{S}[\rho _{y}^{ab}]\leq \sum_{m}p(m|y)%
\mathcal{S}[\rho _{m}^{b}]-\sum_{m}p(m|y)\mathcal{S}[\rho _{m}^{ab}].
\end{equation}%
By adding the previous two inequalities, it follows%
\begin{equation}
\mathcal{I}[\rho _{y}^{ab}]-\sum_{m}p(m|y)\mathcal{I}[\rho _{m}^{ab}]\leq 
\mathcal{S}[\rho _{y}^{ab}]-\sum_{m}p(m|y)\mathcal{S}[\rho _{m}^{ab}],
\end{equation}%
By applying $\sum_{y}p(y)$ to each contribution in the previous inequality,
and using that $\sum_{y}p(m|y)p(y)=p(m),$ lead to%
\begin{eqnarray}
&&\sum_{y}p(y)\mathcal{I}[\rho _{y}^{ab}]-\sum_{m}p(m)\mathcal{I}[\rho
_{m}^{ab}]  \notag \\
&\leq &\sum_{y}p(y)\mathcal{S}[\rho _{y}^{ab}]-\sum_{m}p(m)\mathcal{S}[\rho
_{m}^{ab}],
\end{eqnarray}%
which recovers Eq. (\ref{Mutual_1}).

\section{Bipartite projective measurements}

Here, we apply the main results of Sec. II to the case of bipartite
projective measurements presented in Sec. III.

\subsection{Entropy inequalities}

From Eq. (\ref{entropiasss}), a straightforward calculation gives%
\begin{equation}
\mathcal{S}[\rho _{\Omega }^{ab}]=\mathcal{H}[m]+\sum_{m}p(m)\mathcal{S}%
[\rho _{m}^{a}].  \label{SabProj}
\end{equation}%
From Eq. (\ref{entropias}), the average entropy of the smoothed state reads%
\begin{equation}
\sum_{y}p(y)\mathcal{S}[\rho _{y}^{ab}]=\mathcal{H}[m|y]+\sum_{m}p(m)%
\mathcal{S}[\rho _{m}^{a}],  \label{SabSmooth}
\end{equation}%
where $\mathcal{H}[m|y]=-\sum_{y}p(y)\sum_{m}p(m|y)\ln [p(m|y)]=\mathcal{H}%
[m,y]-\mathcal{H}[y]$ is the conditional entropy of outcomes $\{m\}$ given
outcomes $\{y\}.$ From Eq. (\ref{RhoAB_m}), the average entropy
corresponding to the selective resolution of the non-selective measurement is%
\begin{equation}
\sum_{m}p(m)\mathcal{S}[\rho _{m}^{ab}]=\sum_{m}p(m)\mathcal{S}[\rho
_{m}^{a}].  \label{SabSelective}
\end{equation}%
Using that $0\leq \mathcal{H}[m|y]\leq \mathcal{H}[m]$ \cite{nielsen}, it
follows that the entropy inequalities (\ref{Central}) are fulfilled by the
bipartite system.

From Eqs. (\ref{SabProj}) and (\ref{SabSmooth}), jointly with the inequality
(\ref{HyUpperBound}) it follows%
\begin{equation}
\mathcal{S}[\rho _{\Omega }^{ab}]-\sum_{y}p(y)\mathcal{S}[\rho _{y}^{ab}]=%
\mathcal{H}[m:y]\leq \mathcal{H}[y],  \label{UpperProyBound}
\end{equation}%
where $\mathcal{H}[m:y]=\mathcal{H}[m]-\mathcal{H}[m|y]=\mathcal{H}[m]+%
\mathcal{H}[y]-\mathcal{H}[m,y],$ is the classical mutual information
between the outcomes of both measurements, $\{m\}$ and $\{y\}.$ The
demonstration of the inequality $\mathcal{H}[m:y]\leq \mathcal{H}[y]$ can be
found in \cite{nielsen}. In addition, the inequality (\ref{LowerBound}),
from Eqs. (\ref{SabSmooth}) and (\ref{SabSelective}), reads%
\begin{equation}
\sum_{y}p(y)\mathcal{S}[\rho _{y}^{ab}]-\sum_{m}p(m)\mathcal{S}[\rho
_{m}^{ab}]=\mathcal{H}[m|y]\leq \mathcal{H}[m].  \label{LowerProyBound}
\end{equation}%
The demonstration of the inequality $0\leq \mathcal{H}[m|y]\leq \mathcal{H}%
[m]$ can also be found in \cite{nielsen}. The previous two equations
demonstrate that the general inequalities (\ref{HyUpperBound}) and (\ref%
{LowerBound}) are in fact fulfilled.

In the previous expressions the probabilities read $p(m)=\mathrm{Tr}%
_{a}[\langle m|\rho _{I}|m\rangle ]$ [Eq. (\ref{RhoEme})]. Furthermore, $%
p(y|m)=\mathrm{Tr}_{a}[M_{y}^{\dagger }M_{y}\rho _{m}^{a}]$ [Eq.~(\ref%
{P_y_cond_m})], $p(y,m)=\mathrm{Tr}_{a}[\langle m|\rho _{I}|m\rangle
M_{y}^{\dagger }M_{y}]$ [Eq. (\ref{conjunta})], while the retrodicted
probability $p(m|y)$ [Eq. (\ref{Pmy})] reads $p(m|y)=\mathrm{Tr}_{a}[\langle
m|\rho _{I}|m\rangle M_{y}^{\dagger }M_{y}]/p(y)$ where $p(y)=\sum_{m}%
\mathrm{Tr}_{a}[\langle m|\rho _{I}|m\rangle M_{y}^{\dagger }M_{y}]$ [Eq. (%
\ref{py})].

\textit{Subsystems}: The previous results can also be specified for
subsystem $A$ and $B.$ From Eqs. (\ref{entropiasss}) and (\ref{entropias}),
it follows $\rho _{\Omega }^{b}=\sum_{m}p(m)|m\rangle \langle m|,$ and $\rho
_{y}^{b}=\sum_{y}p(m|y)|m\rangle \langle m|.$ Furthermore, $\rho
_{m}^{b}=|m\rangle \langle m|.$ The inequality Eq. (\ref{Central}),
specified for subsystem $B,$ becomes $\mathcal{S}[\rho _{\Omega }^{b}]=%
\mathcal{H}[m]\geq \sum_{y}p(y)\mathcal{S}[\rho _{y}^{b}]=\mathcal{H}%
[m|y]\geq \sum_{m}p(m)\mathcal{S}[\rho _{m}^{b}]=0,$ because $\mathcal{S}%
[\rho _{m}^{b}]=0.$ Hence, $\mathcal{H}[m]\geq \mathcal{H}[m|y]\geq 0,$
which is a well known inequality valid for Shannon entropies \cite{nielsen}.
Instead for subsystem $A,$ Eq. (\ref{Central}) leads to the non-trivial
relation%
\begin{equation}
\mathcal{S}[\rho _{\Omega }^{a}]\geq \sum_{y}p(y)\mathcal{S}[\rho
_{y}^{a}]\geq \sum_{m}p(m)\mathcal{S}[\rho _{m}^{a}],  \label{EntropyA}
\end{equation}%
where $\rho _{\Omega }^{a}=\sum_{m}p(m)\rho _{m}^{a}$ and $\rho
_{y}^{a}=\sum_{y}p(m|y)\rho _{m}^{a}$ [see Eqs. (\ref{entropiasss}) and (\ref%
{entropias})]. This inequality say us that while the first measurement is
performed over subsystem $B,$ an information gain is also guaranteed for
subsystem $A.$

The inequalities (\ref{HyUpperBound}) and (\ref{LowerBound}) can also be
specified for each subsystem. For subsystem $A$ they become $\mathcal{S}%
[\rho _{\Omega }^{a}]-\sum_{y}p(y)\mathcal{S}[\rho _{y}^{a}]\leq \mathcal{H}%
[y],$ and $\sum_{y}p(y)\mathcal{S}[\rho _{y}^{a}]-\sum_{m}p(m)\mathcal{S}%
[\rho _{m}^{a}]\leq \mathcal{H}[m].$ For subsystem $B$ they lead to the same
classical entropic relations found previously.

\subsection{Mutual information inequalities}

The mutual information under the different measurement schemes are
characterized by Eqs. (\ref{Mutual_2}) and (\ref{Mutual_1}). Each term
appearing in these inequalities is explicitly calculated below.

From the entropy expressions (\ref{SabProj}), (\ref{SabSmooth}), and (\ref%
{SabSelective}), the mutual information associated to the different
measurement stages read%
\begin{equation}
\mathcal{I}[\rho _{\Omega }^{ab}]=\mathcal{S}[\rho _{\Omega
}^{a}]-\sum_{m}p(m)\mathcal{S}[\rho _{m}^{a}],
\end{equation}%
while%
\begin{equation}
\sum_{y}p(y)\mathcal{I}[\rho _{y}^{ab}]=\sum_{y}p(y)\mathcal{S}[\rho
_{y}^{a}]-\sum_{m}p(m)\mathcal{S}[\rho _{m}^{a}].
\end{equation}%
The difference of the previous two equations leads to the lower bound of Eq.
(\ref{positive2}). On the other hand, from Eq.~(\ref{RhoAB_m}) it follows $%
\sum_{m}p(m)\mathcal{I}[\rho _{m}^{ab}]=0,$ which in turn lead to the lower
bound of Eq. (\ref{positive1}).

The upper bounds of Eqs. (\ref{positive2}) and (\ref{positive1}) follows
from the general inequalities (\ref{Mutual_2}) and (\ref{Mutual_1}) written
in terms of Eqs. (\ref{UpperProyBound}) and (\ref{LowerProyBound})
respectively.


\begin{thebibliography}{99}
\bibitem{nielsen} M. A. Nielsen and I. L. Chuang, \textit{Quantum
Computation and Quantum Information} (Cambridge University Press, Cambridge,
England, 2000).

\bibitem{breuerbook} H. P. Breuer and F. Petruccione, \textit{The theory of
open quantum systems}, (Oxford University press, 2002).

\bibitem{aharonov} Y. Aharonov, P. G. Bergman, and J. L. Lebowitz, Time
symmetry in the quantum process of measurement, Phys. Rev. \textbf{134},
B1410 (1964); Y. Aharonov and D. Z. Albert, Is the usual notion of time
evolution adequate for quantum-mechanical systems?, Phys. Rev. D \textbf{29}%
, 223 (1984); Y. Aharonov and D. Z. Albert, Is the usual notion of time
evolution adequate for quantum-mechanical systems? II. Relativistic
considerations, Phys. Rev. D \textbf{29}, 228 (1984).

\bibitem{vaidman} Y. Aharonov and L. Vaidman, Properties of a quantum system
during the time interval between two measurements, Phys. Rev. A \textbf{41}%
,11 (1990); Y. Aharonov and L. Vaidman, Complete description of a quantum
system at a given time, J. Phys. A: Math. Gen. \textbf{24}, 2315 (1991).

\bibitem{molmer} S. Gammelmark, B. Julsgaard, and K. M\o lmer, Past Quantum
States of a Monitored System, Phys. Rev. Lett. \textbf{111}, 160401 (2013).

\bibitem{wiseman} I. Guevara and H. Wiseman, Quantum State Smoothing, Phys.
Rev. Lett. \textbf{115}, 180407 (2015).

\bibitem{tsang} M. Tsang, Time-symmetric quantum theory of smoothing, Phys.
Rev. Lett. \textbf{102}, 250403 (2009).

\bibitem{jaz} A. H. Jazwinski, \textit{Stochastic Processes and Filtering
Theory} (Academic Press, New York, 1970).

\bibitem{recipes} W. H. Press, S. A. Teukolsky, W. T. Vetterling, and B. P.
Flannery, \textit{Numerical Recipes: The Art of Scientific Computing}, 3rd
ed. (Cambridge University Press, New York, 2007).

\bibitem{milburn} H. M. Wiseman and G. J. Milburn, \textit{Quantum
Measurement and Control} (Cambridge University press, 2010).

\bibitem{carmichaelbook} H. J. Carmichael, \textit{An Open Systems Approach
to Quantum Optics}, Lecture Notes in Physics, Vol. M18 (Springer, Berlin,
1993); M. B. Plenio and P. L. Knight, The quantum-jump approach to
dissipative dynamics in quantum optics, Rev. Mod. Phys. \textbf{70}, 101
(1998).

\bibitem{tsanPRA} M. Tsang, Optimal waveform estimation for classical and
quantum systems via time-symmetric smoothing, Phys. Rev. A \textbf{80},
033840 (2009); M. Tsang, Optimal waveform estimation for classical and
quantum systems via time-symmetric smoothing. II. Applications to atomic
magnetometry and Hardy's paradox, Phys. Rev. A \textbf{81}, 013824 (2010);
M. Tsang, H. M. Wiseman, and C. M. Caves, Fundamental quantum limit to
waveform estimation, Phys. Rev. Lett. \textbf{106}, 090401 (2011).

\bibitem{meschede} S. Gammelmark, K. M\o lmer, W. Alt, T. Kampschulte, and
D. Meschede, Hidden Markov model of atomic quantum jump dynamics in an
optically probed cavity, Phys. Rev. A \textbf{89}, 043839 (2014).

\bibitem{murch} D. Tan, S. J. Weber, I. Siddiqi, K. M\o lmer, and K. W.
Murch, Prediction and Retrodiction for a Continuously Monitored
Superconducting Qubit, Phys. Rev. Lett. \textbf{114}, 090403 (2015).

\bibitem{haroche} T. Rybarczyk, B. Peaudecerf, M. Penasa, S. Gerlich, B.
Julsgaard, K. M\o lmer, S. Gleyzes, M. Brune, J. M. Raimond, S. Haroche, and
I. Dotsenko, Forward-backward analysis of the photon-number evolution in a
cavity, Phys. Rev. A \textbf{91}, 062116 (2015).

\bibitem{xu} Q. Xu, E. Greplova, B. Julsgaard, and K. M\o lmer, Correlation
functions and conditioned quantum dynamics in photodetection theory, Phys.
Scr. \textbf{90}, 128004 (2015).

\bibitem{tan} D. Tan, M. Naghiloo, K. M\o lmer, and K. W. Murch, Quantum
smoothing for classical mixtures, Phys. Rev. A \textbf{94}, 050102(R) (2016).

\bibitem{naghi} N. Foroozani, M. Naghiloo, D. Tan, K. M\o lmer and K. W.
Murch, Correlations of the Time Dependent Signal and the State of a
Continuously Monitored Quantum System, Phys. Rev. Lett. \textbf{116}, 110401
(2016).

\bibitem{huard} P. Campagne-Ibarcq, L. Bretheau, E. Flurin, A. Auff\`{e}ves,
F. Mallet, and B. Huard, Observing Interferences between Past and Future
Quantum States in Resonance Fluorescence, Phys. Rev. Lett. \textbf{112},
180402 (2014).

\bibitem{decay} D. Tan, N. Foroozani, M. Naghiloo, A. H. Kiilerich, K. M\o %
lmer, and K. W. Murch, Homodyne monitoring of postselected decay, Phys. Rev.
A \textbf{96}, 022104 (2017).

\bibitem{retro} A. A. Budini, Smoothed quantum-classical states in
time-irreversible hybrid dynamics, Phys. Rev. A \textbf{96}, 032118 (2017).

\bibitem{barnett} T. Pegg and S. M. Barnett, Retrodiction in quantum optics,
J. Opt. B: Quantum and Semiclass. Opt. \textbf{1}, 442 (1999); S. M. Barnett
and D. T. Pegg, Optical state truncation, Phys. Rev. A \textbf{60}, 4965
(1999); S. M. Barnett, D. T. Pegg, and J. Jeffers, Bayes' Theorem and
quantum retrodiction, J. Mod. Opt.\textbf{\ 47}, 1779 (2000); S.\ M.
Barnett, D. T. Pegg, J. Jeffers, and O. Jedrkiewicz, Atomic retrodiction, J.
Phys. B: At. Mol. Opt. Phys. \textbf{33}, 3047 (2000); S. M. Barnett, D. T.
Pegg, J. Jeffers, O. Jedrkiewicz, and R. Loudon, Retrodiction for quantum
optical communications, Phys. Rev. A \textbf{62}, 022313 (2000).

\bibitem{pegg} S. M. Barnett, D. T. Pegg, J. Jeffers, and O. Jedrkiewicz,
Master equation for retrodiction of quantum communication signals, Phys.
Rev. Lett. \textbf{86}, 2455 (2001); D. T. Pegg, S. M. Barnett, and J.
Jeffers, Quantum retrodiction in open systems, Phys. Rev. A \textbf{66},
022106 (2002).

\bibitem{equality} P. Hayden, R. Jozsa, D. Petz, and A. Winter, Structure of
states which satisfy strong subadditivity of quantum entropy with equality,
Comm. Math. Phys. \textbf{246}, 359 (2003); A. Wehrl, General properties of
entropy, Rev. Mod. Phys. \textbf{50}, 221 (1978).

\bibitem{weak} J. Dressel, M. Malik, F. M. Miatto, A. N. Jordan, and R. W.
Boyd, Understanding quantum weak values: Basics and applications, Rev. Mod.
Phys. \textbf{86}, 307 (2014).

\bibitem{sms} A. A. Budini, Quantum jumps and photon statistics in
fluorescent systems coupled to classically fluctuating reservoirs, J. Phys.
B \textbf{43}, 115501 (2010); A. A. Budini, Open quantum system approach to
single-molecule spectroscopy, Phys. Rev. A \textbf{79}, 043804 (2009).

\bibitem{nota1} This occurs, for example, for a qubit system where the
measurements operators are $\{\Omega _{m}\}=\{|\pm \rangle \langle \pm |\}%
\mathrm{\ }$and $\{M_{y}\}=\{|x\pm \rangle \langle x\pm |\},$ where $|\pm
\rangle $ and $|x\pm \rangle $ are the eigenvectors of the $z-$ and $x-$%
Pauli matrixes respectively.

\end{thebibliography}
\end{document}